\documentclass[aps,pra,twocolumn,showpacs,groupedaddress,superscriptaddress,nofootinbib]{revtex4-1}
\usepackage{bm,graphicx,amsmath}
\usepackage[utf8]{inputenc}
\usepackage[T1]{fontenc}
\usepackage{lmodern}
\usepackage{amssymb}
\usepackage{amsmath}
\usepackage{epsfig}
\usepackage{natbib}
\usepackage{color}
\usepackage{pdfpages}
\usepackage{bm}
\usepackage[pdftex,colorlinks,linkcolor=blue,citecolor=blue,urlcolor=blue]{hyperref}
\usepackage{array}
\usepackage{booktabs}
\usepackage[normalem]{ulem}
\usepackage{color}

\newcommand{\pref}[1]{(\ref{#1})}
\newcommand{\eref}[1]{Eq.~\pref{#1}}

\newcommand{\Braket}[1]{\ensuremath{\langle #1 \rangle}}

\newcommand{\comment}[1]{}

\definecolor{gray}{gray}{0.6}

\makeatletter
\AtBeginDocument{\let\LS@rot\@undefined}
\makeatother

\usepackage[absolute]{textpos}

\begin{document}
\title{Damping of Josephson oscillations in strongly correlated one-dimensional atomic gases}
\author{J. Polo}
\affiliation{Univ. Grenoble Alpes, CNRS, LPMMC, F-38000 Grenoble, France} 
\author{V. Ahufinger}
\affiliation{Departament de F\'{\i}sica, Universitat Aut\`{o}noma de Barcelona, E-08193 Bellaterra, Spain} 
\author{F. W. J. Hekking} \thanks{Deceased on May 15, 2017.}
\affiliation{Univ. Grenoble Alpes, CNRS, LPMMC, F-38000 Grenoble, France} 
\author{A. Minguzzi}
\affiliation{Univ. Grenoble Alpes, CNRS, LPMMC, F-38000 Grenoble, France} 
\email[]{Juan.Polo@uab.cat}
\date{\today}
\begin{abstract}
We study Josephson oscillations of two strongly correlated one-dimensional bosonic clouds separated by a localized barrier. Using a quantum-Langevin approach and the exact Tonks-Girardeau solution in the impenetrable-boson limit, we determine the dynamical evolution of the particle-number imbalance, displaying an effective damping of the Josephson oscillations which depends on barrier height, interaction strength and temperature.  We show that the damping originates from the quantum and thermal fluctuations intrinsically present in the strongly correlated gas. Thanks to the density-phase duality of the model, the same results apply to particle-current oscillations in a one-dimensional ring where a weak barrier couples different angular momentum states.
\end{abstract}


\maketitle

The Josephson effect was discovered in 1962 \cite{Josephson_1962} when analyzing the dynamics of two superconductors coupled by a thin layer of insulating material. It is one of the most clear manifestations of macroscopic quantum coherence: its dynamical behavior, based on quantum tunneling, is
fixed by the relative phase between the superconductors and has played a crucial role in the development of technological applications of superconductor materials \cite{barone_1982}.

In ultracold atomic gases, the Josephson effect has been predicted \cite{Javanainen_1986,Shenoy_1997} and experimentally observed in Bose--Einstein condensates trapped in a double-well potential (external Josephson effect \cite{Oberthaler_2005,Pritchard_2005,Steinhauer_2007}) or belonging to two, Raman-coupled, internal states (internal Josephson effect \cite{Holland_1999,Oberthaler_2010}).
Josephson oscillations were also observed in paired atomic Fermi gases \cite{Valtolina1505,Burchianti2018}.
In Bose-Josephson junctions the interplay between tunneling and repulsive interactions gives rise to various dynamical regimes \cite{Shenoy_1997,Leggett_1998,Shenoy_1999}, such as Rabi \cite{Inguscio_2001} and Josephson \cite{Oberthaler_2005,Kruger_2005,Smerzi_2011} oscillations as well as macroscopic quantum self-trapping \cite{Shenoy_1997,Oberthaler_2005}. Weakly-coupled Bose gases are key elements in the development of quantum technologies based on ultracold-atoms, e.g., matter-wave interferometers \cite{Ketterle_1997,Kruger_2005}, sensors \cite{Sidorov_2007}, as well as for quantum computers \cite{Prentiss_2007,Oberthaler_2008} and atomtronics devices \cite{Kwek_2013,Kwek_2014,Amico_2015}.

The theoretical description of Bose-Josephson junctions is generally based on a two-mode model:
at mean field level, a two-mode  Gross-Pitaevskii equation predicts Josephson oscillations as well as macroscopic quantum self trapping \cite{Shenoy_1997,Leggett_1998,Shenoy_1999}. A quantum description based on the two-mode Bose-Hubbard model allows to capture squeezing \cite{Reichel_2010,Polls_2012}, quantum-self trapping \cite{Polls_2010} and the formation of macroscopic superposition states \cite{Hekking_2009}. Theories beyond the two-mode model  show that the latter may provide inaccurate values for the Josephson-plasma frequency \cite{Modugno_2017}, overestimate the coherence \cite{Anglin_2001} as well as the self-trapping effect \cite{Zwerger_2001,Cederbaum_2009,Polkovnikov_2010}, and report collapse and revivals of Josephson oscillations \cite{Walls_1997}.

The Josephson effect becomes particularly intriguing when the quantum character of the Bose gas emerges, beyond the two-mode model description. Low dimensional systems provide an ideal geometry to study the quantum behavior of Bose-Josephson junctions, since quantum fluctuations and correlation effects are enhanced. The strongly-correlated regime for atomic gases trapped in quasi-one-dimensional waveguides has been been reached \cite{Bloch_2004,Weiss_2004} and largely studied experimentally \cite{Bloch_2004,Schmiedmayer_2007,Schmiedmayer_2008,Nagerls_2009,Wei_2017,Schmiedmayer_2017_1}. 

\begin{figure}
\includegraphics[width=1\linewidth]{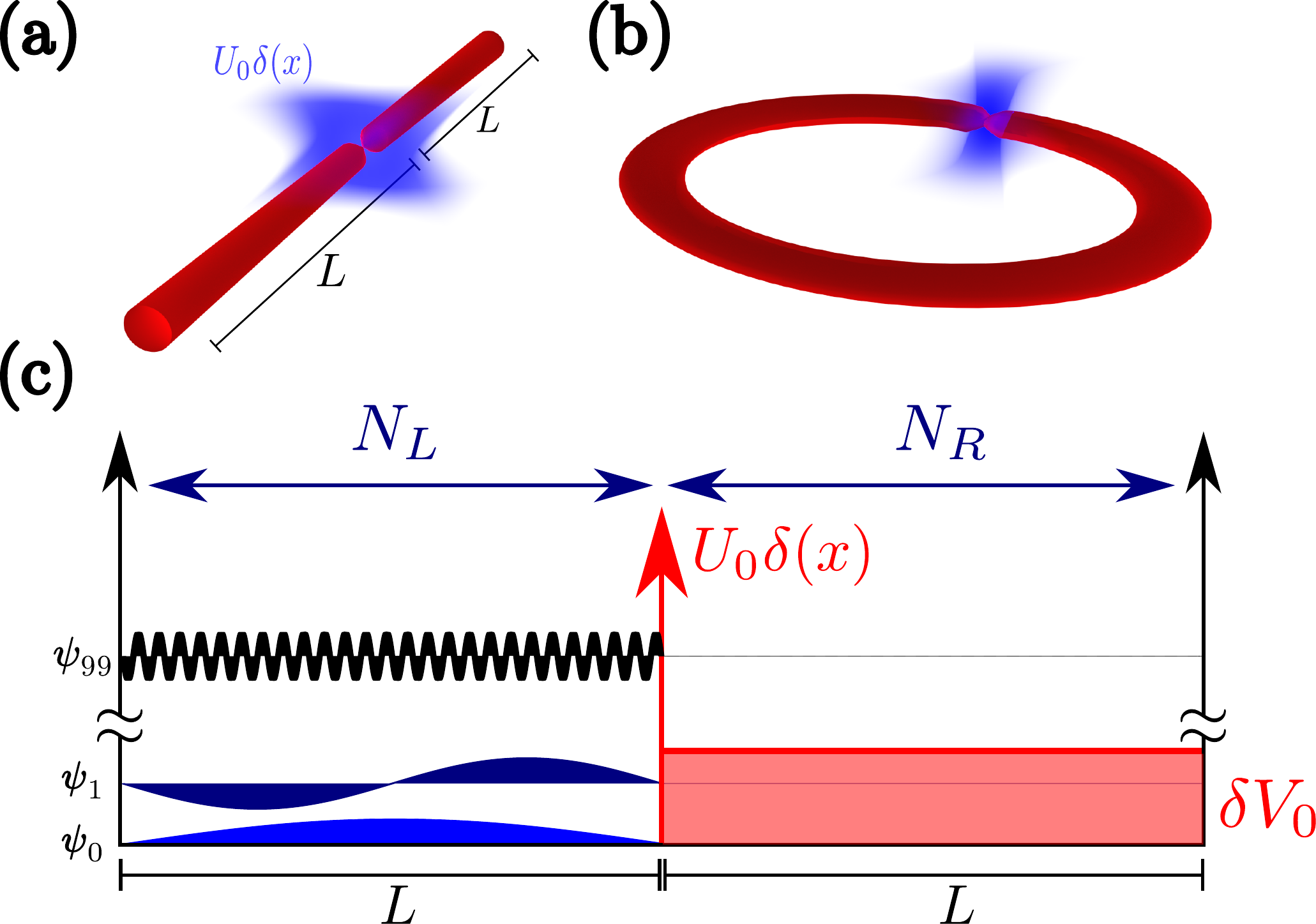}
\caption{Scheme of the geometries considered in this work: two  weakly-coupled atomic waveguides (a) and a ring potential split by a weak barrier (b). (c) Single-particle wavefunctions $\psi_j(x)$  used as initial condition  in the TG solution and the corresponding confining potential.}
\label{fig:fig1}
\end{figure}

In the present work, we focus on the Josephson dynamics among two strongly-correlated one-dimensional bosonic systems coupled head-to-tail through a weak link, as depicted in Fig.~\ref{fig:fig1}~(a). This geometry is complementary to the one of Refs.\cite{Schmiedmayer_2007,Schmiedmayer_2017,Schmiedmayer_2017_1}, where two parallel one-dimensional wires were considered,  and damped Josephson oscillations \cite{Pigneau2018} were observed. In the present case, atom tunnelling between the two wires occurs only through a very small region of both clouds, and, using the Luttinger-liquid (LL) effective theory, we show that the remaining part of the elongated clouds act as effective baths due to their low-energy phonon-like excitations, and provides an effective damping of the Josephson oscillations. An exact solution in the fermionized Tonks-Girardeau limit allows then to obtain the full dynamical behaviour following a quench in the external potential, thus offering an insight on the type of excitations contributing to the oscillations and their damping.

Exploiting the duality of the Luttinger-liquid model, our theoretical framework allows also to describe  bosons in a one-dimensional ring  with a weak barrier under a gauge field, e.g. due to barrier stirring, in which we predict damping in the current oscillations following an initial quench.
Experimental progresses towards the realization of such system have been reported \cite{Campbell_2011,Campbell_2013,Campbell_2014,Perrin_2016,Perrin_2017,Boshier_2009,Boshier_2013,Birkl_2015}, although the one-dimensional regime has not yet been reached.

We start by considering two tunnel-coupled, strongly interacting one-dimensional bosonic fluids, each confined within a tight waveguide of length $L$. To describe the system at intermediate and large interactions, we use the Luttinger-liquid low-energy theory, corresponding to a quantum hydrodynamic theory for density and phase fluctuations (see e.g.~\cite{Cazalilla_2004}).
The total Hamiltonian is given by two Luttinger liquids, $\hat{H}_{LL\pm}$, with $+$ and $-$ for  the right and left waveguide respectively, coupled by a tunnel term,  yielding  a special limit of a boundary sine-Gordon model (see \cite{Saleur_1996} and refs. therein):
\begin{align}
\!\!\!\hat{H}_{LL\pm}&=\frac{\hbar v_{\pm} K_{\pm}}{2\pi}\!\!\!\int_0^{L}\!\!\!dx\bigg[(\partial_x \hat{\varphi}_\pm(x))^2 +\frac{1}{K_{\pm}^2}(\partial_x\hat{\theta}_\pm(x))^2\bigg]\!\!\!
\label{eq:LLplusminus}
\\
\hat{H}_t&=-E_J\cos[\hat{\varphi}_{+}(0^+)-\hat{\varphi}_{-}(0^-)],
\label{eq:totalhamiltniansum}
\end{align}
where   $\partial_x\hat{\theta}_\pm(x)/\pi$ is the density-fluctuation field operator, conjugate to the phase operator $\hat{\varphi}_\pm(x)$, fulfilling $[\partial_{x}\hat{\theta}_\pm(x)/\pi,\hat{\varphi}_\pm(x')]=-i\delta(x-x')$ \cite{Fisher_1992}.
The LL Hamiltonians (\ref{eq:LLplusminus}) are expressed in terms of two parameters, the velocities $v_{\pm}$ of the low-energy excitations, and the dimensionless Luttinger parameters $K_{\pm}$, related to the compressibility of each cloud \cite{Cazalilla_2004}. In the following we shall assume for simplicity that the two atomic waveguides are identical and  set $v_{+}=v_-=v$ and $K_+=K_-=K$.
The tunnel Hamiltonian $\hat{H}_t$ describes the presence of a large, localized barrier whose microscopic parameters determine the Josephson energy $E_J$ (see e.g. \cite{Weiss_1996}). 

We proceed by representing the Hamiltonians (\ref{eq:LLplusminus}) and (\ref{eq:totalhamiltniansum}) on the normal modes basis of each Luttinger liquid, namely the zero modes $\hat{N}_{\pm}$, counting the particle number in each waveguide, their conjugates phases $\hat \phi_{0\pm}$, as well as the position $ \hat{Q}_{\mu\pm}$ and momentum $\hat{P}_{\mu\pm }$ operators for each excitation with wavevector $k_\mu= \pi \mu/L$ and frequency $\Omega_\mu=vk_\mu$. We then focus on the relative-variable problem, which is non-quadratic due to the tunnel barrier term (\ref{eq:totalhamiltniansum}), and we introduce $\hat{N}\equiv \frac{1}{2}(\hat{N}_{+}-\hat{N}_{-})$, $\hat{\phi}_0\equiv\hat{\phi}_{0+}-\hat{\phi}_{0-}$ for the zero modes, and $\hat{Q}_\mu\equiv \hat{Q}_{\mu+}-\hat{Q}_{\mu-}$ and $\hat{P}_\mu\equiv \frac{1}{2}(\hat{P}_{\mu+}-\hat{P}_{\mu-})$ for the excited modes. The resulting Hamiltonian reads \cite{suppl}:

\begin{align}
\hat{H}^{rel}_{T}&=
\frac{\hbar^2}{2ML^2}(\hat{N}-N_{\text{ex}})^2
-E_J\cos\left(\hat{\phi}_{0}\right)
\label{eq:Hamiltonian_trans}\\&
+\sum_{\mu\ge1}\left[
\frac{1}{2M}\left( \hat{P}_{\mu} +\frac{\sqrt{2}\hbar}{L}(\hat{N}-N_{\text{ex}})\right)^2
+\frac{1}{2}M\Omega_{\mu}^2\hat{Q}_{\mu}^2
\right]
\nonumber
\end{align}%
with effective mass $M=\hbar K/2\pi vL=K^2m/2\pi^2N_0$, $N_0$ being the average particle number in each tube and $N_{\text{ex}}\ll N_0$ is the excitation imbalance, which may be tuned by a suitable choice of the initial conditions. We identify in Eq.~(\ref{eq:Hamiltonian_trans}) a {\it quantum particle} term corresponding to the two collective variables $\hat{N}$ and $\hat{\phi}_0$, a bath of harmonic oscillators formed by the excited modes, and a coupling term  $\propto \hat{P}_{\mu}\hat{N}$, obtained by expanding the second line of Eq.(\ref{eq:Hamiltonian_trans}). The same structure is found in the Caldeira-Leggett Hamiltonian \cite{Leggett_1981,Leggett_1983,SCHON_1990}, however, in our model, the bath of harmonic oscillators is intrinsic in the model, originated from the phonon excitations in the Bose fluid, while in the Caldeira-Leggett model it is phenomenologically introduced.  The first line of Eq.~(\ref{eq:Hamiltonian_trans}) corresponds to the familiar Josephson Hamiltonian, where two regimes are possible depending on the ratio of the Josephson  $E_J$ and kinetic $E_Q=\hbar^2/M L^2= 2\Delta E/K$ energies,  with $\Delta E=\hbar\pi v/L$ being the level spacing among phonon modes of the bath. Notice that  $E_J$ and $E_Q$ depend on interactions, since the tunnel energy is renormalized by quantum fluctuations \cite{Minguzzi_2014}, and both the sound velocity and the Luttinger parameter vary with  interaction strength \cite{Cazalilla_2004}.

We start from the case $E_J\gg E_Q$,  where the Josephson potential term $- E_J \cos(\hat{\phi}_0)$  dominates upon the kinetic energy. Starting from an initial particle imbalance among the two wires, its   dynamical evolution is readily obtained from the Heisenberg equations of motion, and takes a quantum Langevin form: %
{\allowdisplaybreaks
\begin{equation}
\ddot{\hat{N}}+\omega_0^2\cos(\hat{\phi}_0)\hat{N}+\int_{0}^{t}dt'\,\gamma_{N}(t,t')\dot{\hat{N}}(t') =\xi_{N}(t)\label{eq:eq_motion_n}
\end{equation}%
with   \cite{suppl}  ${\omega_0=\sqrt{E_JE_Q}/\hbar}$  the Josephson frequency,
$\gamma_{N}(t,t')$
  the memory-friction kernel,  and $\xi_{N}(t)$
    the quantum noise generated by the phonon bath.
    $\gamma_{N}(t,t')$ can be approximated to be local in time in the case of long wires where many excited phonons contribute to the bath and in the low-energy regime, where the high-energy cutoff of the LL theory is the largest energy scale in the problem. In the high-temperature regime we have  $\langle\xi_{N}(t)\rangle=0 $ and $\langle\xi_{N}(t)\xi_N(t')\rangle=\eta \delta(t-t')$, with $\eta\!=\!\!2E_J^2k_BT\!/\hbar^2\! MLv$. 

For small phase oscillations,  the average relative number $N_{\text{LL}}=\Braket{\hat{N}}$ in Eq.~(\ref{eq:eq_motion_n}) is then described by a damped harmonic oscillator with frequency $\omega_J=\sqrt{\omega_0^2-\gamma^2}$ and  damping rate $\gamma=\pi E_J/\hbar K$ \cite{suppl}. In the weakly interacting limit, where $K\sim 1/\sqrt{g_{1D}}$ and $v_s\sim \sqrt{g_{1D}}$ with $g_{1D}$ the 1D interaction strength,  we recover  the predictions of the two-mode model in its small-oscillation limit, i.e. we find  $E_Q\propto g_{1D}$ and $\gamma/E_Q$ vanishing for $g_{1D}\rightarrow 0$, yielding undamped Josephson oscillations.  At increasing interactions $E_Q$ increases, being related to the compressibility of the system,   and $E_J$ decreases, since it is renormalized by larger and larger phase fluctuations. Since $\gamma_Q\equiv\gamma/\omega_0 = \pi\sqrt{E_J}/\sqrt{E_Q}K$, we predict that Josephson oscillations will be more and more damped at increasing interactions. Hence, quite interestingly, while remaining in the regime  $E_J\gg E_Q$, both the underdamped and the overdamped Josephson oscillations can be accessed. Using realistic experimental values  \cite{Schmiedmayer_2007,Schmiedmayer_2008}, i.e. $E_J/\hbar=2\pi\times80,\,200,\,900$ Hz, $L\approx 6.8$ $\mu$m, $v \approx 6.7$ mm/s and a 1D interaction strength $g_{1D}/\hbar=0.84$ mm/s which leads to $E_Q/\hbar\approx 245$ Hz for  $N= 20$, we estimate $\omega_0\approx2\pi\!\times\!(50-160)$ Hz and $\gamma\approx 0.4-20$ $\omega_0$. Notice also that at fixed interactions one may explore the crossover from underdamped to overdamped oscillations  by tuning the barrier strength.

In Fig.~\ref{fig:fig2} we show the damped Josephson oscillations of the relative number between the two clouds, at varying barrier and interaction strength. The noise in Eq.~(\ref{eq:eq_motion_n}) yields stochastic fluctuations in the dynamics \cite{suppl}, indicated as shaded areas in the figure. At long times $t\gg 1/\gamma$ and in the high-temperature regime we have $\Delta N_{\text{LL}}=\langle (\hat{N}-\langle \hat{N}\rangle)^2 \rangle^{1/2}=\sqrt{\frac{ML^2}{\hbar^2}k_BT}$, which coincides with the high-temperature limit found using the fluctuation dissipation theorem \cite{suppl,Weiss_2008}. Of course, in any closed, finite quantum system revivals are expected, and would occur if a discrete phonon spectrum is used. In a semi-classical approach, for exemple, revivals can be viewed as a resyncronization of the bath modes \cite{Nagaev_2002}.

\begin{figure}
\includegraphics[width=0.65\linewidth]{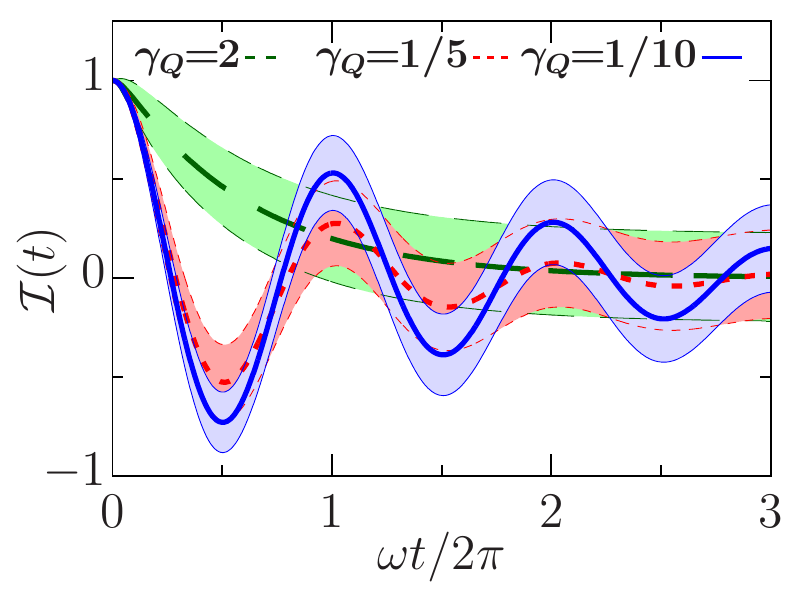}
\caption{(Color online)
Dynamics of imbalance ${\cal I} (t)$ (dimensionless) for relative number $N(t)/N(0)$ in tunnel-coupled wires,  or for  current oscillations $J(t)/J(0)$ in a ring,  from the LL approach, for various values of $\gamma_Q\!=\!\gamma/ \omega_0$ or $\gamma_Q\!=\!\gamma^{\textrm{r}}/ \omega^{\textrm{r}}_0$, respectively. The corresponding uncertainties due to the stochastic noise are indicated in shaded areas.}
\label{fig:fig2}
\end{figure}

In the opposite regime $E_J\ll E_Q$, the phase is only weakly pinned and displays large fluctuations. The dynamics is most suitably described in the Fock basis for the relative number. The energy levels of the quantum particle seen in \eref{eq:Hamiltonian_trans} can be described as a function of the number of excitations $N_{\text{ex}}$, which plays the role of quasi-momentum in crystals, and takes the form of a sequence of parabolas $\varepsilon_n(N_{\text{ex}})=E_Q(n-N_{\text{ex}})^2/2$, with $\hat N|n\rangle=n |n\rangle$, with gaps of amplitude $E_J$ opening at semi-integer values of $N_{\text{ex}}$. Close to the anticrossing points $N_{\text{ex}}=\pm 1/2, \pm 3/2,...$ the system behaves as an effective two-level model. In this case the Josephson dynamics correspond to the Rabi oscillations of the quantum particle, with frequency $E_J/\hbar$. Due to the large value of $E_Q$, which also fixes the scale of bath-modes level-spacing, in this regime there is no effect of the bath modes on the quantum particle. 

The Luttinger liquid model is very useful because it allows to describe a large range of interaction strengths, though, it remains an effective model. In the following we take a complementary approach and solve the exact quantum mechanical evolution in the limit of infinitely strong repulsive interactions, i.e, the Tonks-Girardeau (TG) regime \cite{Girardeau_1960}, corresponding to the case $K=1$ of the LL theory.
In this limit, using the time-dependent Bose-Fermi mapping \cite{Girardeau_1960,Wright_2000,Girardeau_2005}, the many-body wavefunction $\Psi_{TG}$ can be written  as
\begin{equation}
\Psi_{TG}(x_1,...,x_N)=\Pi_{1\le j\le \ell\le N}\text{sgn}(x_j-x_\ell)\, \text{det}[\psi_{k}(x_j,t)],
\end{equation}
where $\psi_j(x,t)$ is the solution of the single-particle 
Schr\"odinger equation $ i \hbar \partial_t \psi_j= (-\hbar^2 \partial_x^2/2m + V(x,t))\psi_j$ with initial conditions $\psi_j(x,0)=\psi_j$ being the eigenfunctions of the Schr\"odinger problem at initial time. To induce Josephson oscillations, we perform a quench in the confining potential $V(x,t)$ \cite{Minguzzi_2015,Hekking_2011}, taken as a box potential separated in two parts by a delta barrier $U_0 \delta(x)$ with an imbalance $\delta V_0$ between left and right waveguide at initial time (see Fig.~\ref{fig:fig1}~(c)). $\delta V_0$ is then set to zero during the time evolution, inducing $2N_{\text{ex}}$ excitations above the Fermi energy.
 The total density profile   of the TG gas at finite temperature is $n(x,t)=\sum_{j=0}^{\infty}f(\epsilon_j)|\psi_j(x,t)|^2$   \cite{Millard_1969,Wright_2002},  with $f(\epsilon_n)$ the Fermi-Dirac distribution  and $\epsilon_{n}$ the $n$th single-particle eigenenergy, allowing us to obtain the relative particle number  $N(t)=N_L-N_R$, with $N_L=\int_{-L}^0\!\!dx\, n(x,t)$ and $N_R=\int_{0}^L\!\!dx\, n(x,t)$.

\begin{figure}
  \includegraphics[width=1\linewidth]{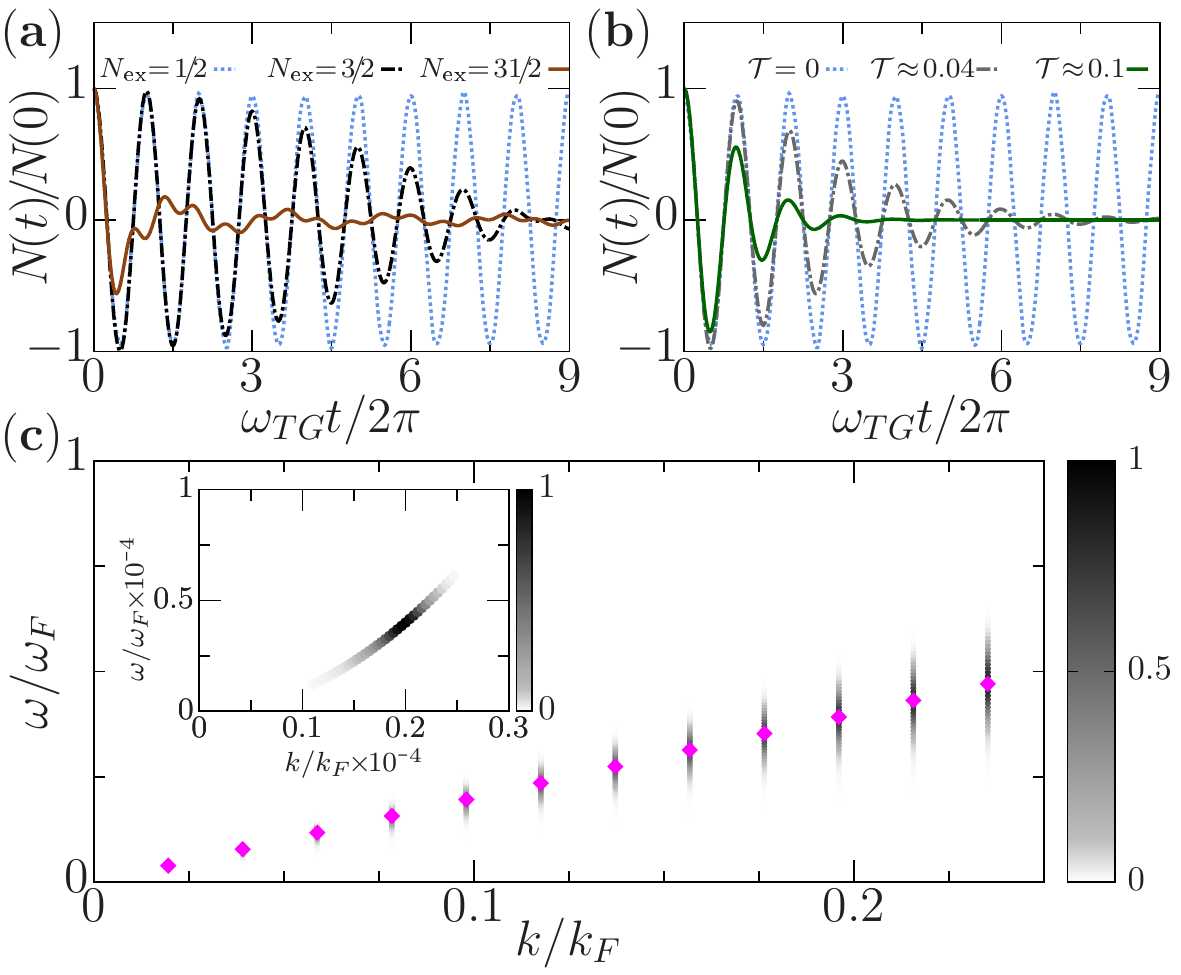}
\caption{(Color online)
Relative-number oscillations in
the TG regime following a quench of the initial step potential $\delta V_0$: (a) at zero temperature for  $\delta V_0/E_F = 0.07$ (blue-dotted line), $0.14$ (black dashed line) and $0.72$ (brown solid line), with $E_F$ the Fermi energy; (b) at finite temperature for $\delta V_0/E_F = 0.07$.
(c) exact TG excitation spectrum (dimensionless, grey-scale points) in the frequency - wavevector plane for  $T/T_F=0.1$, and LL excitation spectrum (magenta points); inset: zoom on the small-$k$ region.
 In all panels  $N_T=101$ and  $\lambda \equiv 2mLU_0/\hbar^2\pi^2N_T  = 200$. 
}
\label{fig:fig3}
\end{figure}

In Fig.~\ref{fig:fig3}~(a) we show the exact dynamics of $N(t)$
following the quench in the step potential.
We observe that for an initial small imbalance,  corresponding to $N_{\text{ex}}=1/2$, undamped oscillations occur, with frequency $\omega_{TG}=\epsilon_{N+1}-\epsilon_N$. For a larger imbalance an effective damping appears,  as a consequence of the several frequencies associated to the  excitations involved in the dynamics. 
The exact solution allows also to address the long-time dynamics  where oscillations display revivals  \cite{suppl} since the system has finite size.
In order to make connection with the LL model, we notice that for bosons in the TG regime   $E_J=\hbar \omega_{TG}$ and $E_Q=\hbar^2 \pi^2 N/mL^2$. For the parameters used in  Fig.~\ref{fig:fig3}  $E_J/E_Q= 4\times 10^{-3}$. Hence,  the oscillations observed in the exact solutions at small $\delta V_0$ are the undamped  Rabi oscillations of the quantum particle predicted by  the LL model. For larger imbalance, the exact dynamics corresponds to large-amplitude oscillations, beyond the LL treatment.

Figure~\ref{fig:fig3}~(b) shows the small-imbalance dynamics at finite temperature. At difference from the predictions of the LL model, we find damped oscillations. In order to pinpoint the origin of this damping, using the exact solution we calculate the spectral function for the system at finite temperature, see  Fig.~\ref{fig:fig3}(c). While the exact spectral function contains multiple particle-hole excitations,
the LL model assumes a linear excitation spectrum. This is an excellent approximation at low energy, and in particular for the energy scales involved in the dynamics of the current study. The exact spectral function contains also  several low-energy excitations with frequencies of order  $E_J$  (inset of  Fig.~\ref{fig:fig3}(c)), which are associated to the presence of a finite barrier and  give rise to the observed  damping. These modes  are absent in the infinite-barrier case corresponding  to the LL Hamiltonian (\ref{eq:LLplusminus}). As a main conclusion of this analysis, the exact solution validates the frequency of the Josephson oscillations predicted in the LL model, and the fact that oscillations may be  damped by an intrinsic bath, made of low-energy excitations.

The Luttinger-liquid  analysis applies also to a dual system, made of ultracold bosons confined in a ring trap of circumference length $L$, containing a small, localized barrier and subjected to an artificial gauge field $\Omega$. In this system, we follow the dynamical evolution of the average current as a function of time, following a sudden quench of $\Omega$. This can be induced, for instance, by transferring orbital angular momentum   on the atoms with a Laguerre-Gauss beam \cite{Phillips_2007}, by phase imprinting \cite{Hadzibabic_2012}, by stirring a potential barrier \cite{Clark_2013} or by modulating an artificial gauge field \cite{Patrik_2011}. We model the system by a single LL Hamiltonian that describes the particles in the ring,
\begin{equation}
\hat{H}_{LL}=\frac{\hbar v K}{2\pi}\!\!\!\int_0^{L}\!\!\!dx\big[(\partial_x \hat{\varphi}(x)-\frac{2\pi}{L}\Omega)^2 +\frac{1}{K^2}(\partial_x\hat{\theta}(x))^2\big],%
\end{equation}
plus a weak delta potential barrier $U(x)=U_0\delta(x)$, with corresponding Hamiltonian $H_b=2n_0U_{\text{eff}}\cos(2\hat{\theta} (x=0))$, with $n_0=N_T/L$ and 
  $U_{\text{eff}}$ the effective barrier strength \cite{Minguzzi_2014}. Notice that the duality of the model follows from the density-phase duality of the LL Hamiltonian as well as the duality between strong and weak-barrier limits in the LL description. 
In the ring geometry, the relevant collective variables are the current and zero-mode density field operator, fulfilling $[\hat{\theta}_0,\hat{J}]=i/2$.
By following a procedure similar to the  coupled waveguide case \cite{suppl}, we find the effective Hamiltonian: 
\begin{align}
&\hat{H}_T=\frac{\hbar^2(2\pi)^2}{2M_\text{r}L^2}\left(\hat{J}-\Omega\right)^2 +  2n_0 U_{\text{eff}}\cos(2\hat{\theta}_0) \label{eq:hamiltonianring}\\
+& \sum_{\mu\ge1}\Bigg[ \frac{1}{2M_\text{r}} \left( \hat{P}_\mu + \frac{4\pi\sqrt{2}\hbar}{L}\left(\hat{J}-\Omega\right) \right)^2
+ \frac{1}{2}M_\text{r}\Omega_\mu^2\hat{Q}_\mu^2  \bigg] \nonumber
\end{align}
where the quantum particle is now the current,  $M_\text{r}=\frac{\hbar\pi}{vLK}$ and $\Omega_\mu=vk_\mu$ are the  mass and frequencies of the bath modes, and in this case $E_Q^{\text{r}}=\hbar^2(2\pi)^2/M_\text{r}L^2$ and $E_J^{\text{r}}=2n_0 U_{\text{eff}}$.
When $E_J^{\text{r}}> E_Q^{\text{r}}$ the small oscillations of the current   (see Fig.~\ref{fig:fig2}) are again described by a harmonic oscillator with frequency $\omega^{\text{r}}_0=\sqrt{E^{\text{r}}_QE^{\text{r}}_J}/\hbar$ and damping rate $\gamma^{\text{r}}=4\pi E^{\text{r}}_JK/\hbar$. The effective damping originates from the  phonon modes of the ring. Notice that in this dual model damping decreases at increasing interactions. In the opposite regime $E_J^{\text{r}}< E_Q^{\text{r}}$, for small imbalances  we expect undamped Rabi oscillations among angular momentum states. The exact TG solution for a quantum quench  of the artificial gauge field on a ring shows  weakly damped oscillations and the formation of non-classical states \cite{Hekking_2011}. 

In conclusion, by combining Luttinger-liquid theory and an exact solution at infinite interactions we have studied the Josephson oscillations of particle imbalance among two atomic waveguides as well as particle-current oscillations along a ring. In both cases, we have found that an intrinsic damping is present in the oscillations due to the coupling with the collective excitations in the system.
Our approach also yields analytical expressions for the natural frequencies and damping rates as a function of the microscopic parameters of the model. In a similar fashion, the bath phonon modes  gives rise to damping of current-current time correlation functions \cite{Prokof_2017}.
Our results are relevant not only to ongoing studies on the bosonic Josephson effect at different interactions strengths, but also to future developments of quantum devices in which dissipation and thermalization can be limiting factors to perform quantum computations. Moreover, the results in ring potentials are particularly relevant to current experiments, in particular to Atomtronic devices \cite{Holland_2009,Campbell_2014,Amico_2015} where the interplay of interactions and barrier strength is crucial when creating persistent currents.

We would like to thank R. Dubessy, M. Filippone, P. Pedri and H. Perrin for useful discussions. The authors acknowledge financial support through the ANR project SuperRing (ANR-15-CE30-0012-02). J.P and V.A acknowledge the Spanish Ministry of Economy and Competitiveness (MINECO) (Contract No. FIS2014-57460-P) and J.P. acknowledges the mobility grant EEBB-I-15-10194.

\bibliographystyle{apsrev4-1}

\bibliography{references}
\clearpage



\section*{Supplementary material}

# Supplemental Material for:
# Damping of Josephson oscillations in strongly correlated one-dimensional atomic gases

J. Polo,[1] V. Ahufinger,[2] F. W. J. Hekking,[1, *] and A. Minguzzi[1]

[1] *Univ. Grenoble Alpes, CNRS, LPMMC, F-38000 Grenoble, France*
[2] *Departament de Física, Universitat Autònoma de Barcelona, E-08193 Bellaterra, Spain*

(Dated: July 17, 2018)

## CONTENTS

## I. MODE EXPANSION AND EFFECTIVE HAMILTONIAN

We outline the derivation yielding to the effective Hamiltonian for the two 1D finite wires coupled through a weak link, Eq. (3) of the main text, and to the ring potential separated by a weak barrier, Eq. (6) of the main text.

### A. Weakly-coupled atomic waveguides

The effective Hamiltonian shown in Eq. (3) of the main text is obtained by introducing the following mode expansion:

$$\hat{\phi}_\pm(x) = \hat{\phi}_{0\pm} + \frac{1}{\sqrt{L}} \sum_{\mu \geq 1} \Phi_\mu(x) \hat{Q}_{\mu\pm},$$

$$\hat{n}_\pm(x) = \frac{\hat{N}_\pm}{L} + \frac{\sqrt{L}}{\hbar} \sum_{\mu \geq 1} \Phi_\mu(x) \hat{P}_{\mu\pm}, \quad \text{(S1)}$$

into the total Hamiltonian (Eq. (1) and Eq. (2) of the main text). In Eq. (S1), $\hat{n}_\pm(x) \equiv \frac{\partial_x \hat{\theta}_\pm(x)}{\pi}$ is the fluctuation of the density field operator and $\hat{N}_\pm$ is the particle-number operator. The commutation relation $[\frac{\partial_x \hat{\theta}_\pm(x)}{\pi}, \hat{\phi}_\pm(x')] = -i\delta(x-x')$ yields $[\hat{N}_\pm, \hat{\phi}_{0\pm}] = -i$ and $[\hat{P}_{\mu\pm}, \hat{Q}_{\mu'\pm}] = -i\hbar\delta_{\mu,\mu'}$ provided that the mode expansion forms a complete orthonormal basis, i.e., $\sum_{\mu=0} \Phi_\mu(x)\Phi_\mu(x') = \delta(x-x')$ and $\int_0^L dx\, \Phi_\mu(x)\Phi_{\mu'}(x) = \delta_{\mu,\mu'}$. Moreover, in order to diagonalize the Hamiltonian, the mode basis must fulfill $\partial_x^2 \Phi_\mu = -k_\mu^2 \Phi_\mu$. Imposing open boundary conditions, i.e., $\partial_x \Phi_\mu(x)|_{x=0} = \partial_x \Phi_\mu(x)|_{x=L} = \partial_x \Phi_\mu(x)|_{x=-L} = 0$, yields $k_\mu = \pi\mu/L$; with $\mu \geq 1$ taking positive integer values and $\Phi_\mu(x) = \sqrt{2/L}\cos(k_\mu x)$. A shift operator $U = e^{i\frac{\sqrt{2}}{L}\hat{N}\sum_{\mu\geq 1}\hat{Q}_\mu}$ is then applied to the total Hamiltonian in order to remove the non-zero modes contribution from the tunneling term (Eq. (2) of the main text). Finally, by using the relative coordinates introduced in the main text together with the definition of the effective mass and frequency of the non-zero phonon modes, the effective Hamiltonian given in Eq. (3) of the main text is found.

---

* Deceased on May 15, 2017.

## B. Ring trap with weak barrier

In the ring configuration we use a single Luttinger liquid (LL) Hamiltonian plus a weak barrier that creates a small density notch in the system. The periodic boundary conditions describing the ring geometry lead to the following relations between the phase and density field operators $\hat{\phi}(x+L) = \hat{\phi}(x) + 2\pi(\hat{J} - \Omega)$ and $\hat{\Theta}(x+L) = \hat{\Theta}(x) + \pi \hat{N}$, where $L$ is the ring circumference, $\Omega$ is the artificial gauge field, and $\hat{J}$ and $\hat{N}$ are the angular momentum and particle-number operators. Note that the operator $\hat{\Theta}(x)$ is related to the density field operator, $\hat{n}(x) = \partial_x \hat{\theta}(x)/\pi$, by $\hat{\Theta}(x) = \hat{\theta}(x) + N_0 \pi x / L$ [52]. The commutation relation among density and phase field operators can be rewritten as $[\frac{\hat{\theta}(x)}{\pi}, \partial_{x'}\hat{\phi}(x')] = i\delta(x-x')$; thus, we can regard $\partial_x \hat{\phi}(x)$ as the momentum conjugate to $\hat{\theta}(x)$. The mode expansion used in the ring configuration is therefore given by:

$$\hat{\theta}(x) = \hat{\theta}_0 + \frac{\pi}{\sqrt{L}} \sum_{\mu \geq 1} \Phi_\mu(x) \hat{Q}_\mu, \tag{S2}$$

$$\partial_x \hat{\phi}(x) = \frac{2\pi}{L}(\hat{J} - \Omega) + \frac{\sqrt{L}}{\hbar} \sum_{\mu \geq 1} \Phi_\mu(x) \hat{P}_\mu, \tag{S3}$$

where $[\hat{Q}_\mu, \hat{P}_\mu] = i\hbar$ and $[\hat{\theta}_0, \hat{J}] = i/2$, with $\hat{\theta}_0$ and $\hat{J}$ being the zero mode density fluctuation operator, and the angular momentum operator respectively. Again, it is assumed that the mode expansion functions $\Phi_\mu(x)$ form a complete orthonormal basis, i.e. $\sum_{\mu \geq 0} \Phi_\mu^*(x) \Phi_\mu(x') = \delta(x-x')$ and $\int_0^L dx\, \Phi_\mu(x)^* \Phi_{\mu'}(x) = \delta_{\mu,\mu'}$. The diagonalization of the LL ring Hamiltonian is achieved by assuming $\partial_x^2 \Phi_\mu(x) = -k_\mu^2 \Phi_\mu(x)$. Moreover, the boundary conditions of our system, $\partial_x \Phi_\mu(x)|_{x=0} = \partial_x \Phi_\mu(x)|_{x=L}$ and $\Phi_\mu(0) = \Phi_\mu(L)$ lead to $\Phi_\mu(x) = \sqrt{\frac{1}{L}} \exp(ik_\mu x)$, with $k_\mu = 2\pi\mu/L$ and where $\mu$ takes all integer values. Finally, as in Sec. IA of the supplemental material, we apply a shift operator, $\hat{U} = e^{i \frac{2\pi\sqrt{2}}{L} \hat{J} \sum_{\mu \geq 1} \hat{Q}_\mu}$, to shift the non-zero modes from the nonlinear term originating from the barrier to obtain the effective Hamiltonian given in Eq. (6) of the main text.

## II. EQUATIONS OF MOTION FOR COUPLED WIRES

The equations of motion for the relative particle-number and relative phase in the Luttinger liquid theory are obtained using the Heisenberg equation, $\frac{d}{dt}\hat{A} = \frac{i}{\hbar}[\hat{A}, \hat{H}]$. This yields the following quantum Langevin equations of motion for the quantum particle degrees of freedom:

$$\ddot{\hat{N}} + \int_0^t dt'\, \gamma_N(t,t') \dot{\hat{N}}(t') + \omega_0^2 \cos(\hat{\phi}_0) \hat{N} = \xi_N(t), \tag{S4}$$

$$\ddot{\hat{\phi}}_0 + \int_0^t dt'\, \gamma_{\phi_0}(t,t') \dot{\hat{\phi}}_0(t') + \omega_0^2 \sin(\hat{\phi}_0) = \xi_{\phi_0}(t), \tag{S5}$$

as well as for the bath:

$$\ddot{\hat{Q}}_\mu + \Omega_\mu^2 \hat{Q}_\mu = -\sqrt{2} \omega_0^2 L \sin(\hat{\phi}_0). \tag{S6}$$

In the above equations we have $\gamma_N(t,t') = \cos(\hat{\phi}_0(t)) 2\omega_0^2 \sum_{\mu \geq 1} \cos(\Omega_\mu(t-t'))$, $\gamma_{\phi_0}(t,t') = \cos(\hat{\phi}_0(t')) 2\omega_0^2 \sum_{\mu \geq 1} \cos(\Omega_\mu(t-t'))$, $\xi_N(t) = -\cos(\hat{\phi}_0) \frac{\sqrt{2} E_J}{\hbar L} \sum_{\mu \geq 1} \dot{\hat{q}}_\mu$ and $\xi_{\phi_0}(t) = -\frac{\sqrt{2}}{L} \sum_{\mu \geq 1} \Omega_\mu^2 \hat{q}_\mu$ and we have defined $\omega_0 = \sqrt{E_J/ML^2} = \sqrt{E_J E_Q}/\hbar$ with $E_Q = \hbar^2/ML^2$. The solution of Eq. (S6) is:

$$\hat{Q}_\mu = \hat{q}_\mu - \sqrt{2}\omega_0^2 L \int_0^t dt'\, G_{Q,\mu}(t,t') \sin(\hat{\phi}_0(t')), \tag{S7}$$

where $G_{Q,\mu}(t,t') = \frac{1}{\Omega_\mu} \sin(\Omega_\mu(t-t')) H(t-t')$ with $H(t-t')$ being the Heaviside step function, and where $\hat{q}_\mu$ is the homogeneous solution of the bath mode $\mu$, obtained setting to zero the coupling to the quantum particle.

For small phase oscillations, i.e. $\cos(\hat{\phi}_0) \approx 1$ and $\sin(\hat{\phi}_0) \approx \hat{\phi}_0$, the relative particle-number equation of motion in dimensionless units with respect to the natural frequency $\omega_0$ reads:

$$\ddot{\hat{N}} + 2\gamma_Q \dot{\hat{N}} + \hat{N} = \chi_N(\tau), \tag{S8}$$

where $\gamma_Q = \gamma/\omega_0$, with $\gamma = E_J/2MLv = \pi E_J/\hbar K$ being the intrinsic damping rate due to the coupling to phonons, and $\chi_N(\tau) = \xi_N(\tau/\omega_0)/\omega_0^2$ being the rescaled noise operator. The solution of Eq. (S8) follows from linear-response theory, and reads

$$\hat{N}(\tau) = \hat{n}_0(\tau) + \int_0^\tau d\tau' \, G_N(\tau,\tau')\chi_N(\tau'), \tag{S9}$$

with $\hat{n}_0(\tau)$ being the homogeneous solution, and where the retarded Green's function reads:

$$G_N(\tau,\tau') = \frac{1}{\Omega}e^{-\gamma_Q(\tau-\tau')}\sin(\Omega(\tau-\tau'))H(\tau-\tau'), \tag{S10}$$

with $\Omega = \sqrt{1-\gamma_Q^2}$ corresponding to the oscillation frequency $\omega_J = \Omega\omega_0$ in the main text. The homogeneous solution for the average particle-number of the damped harmonic oscillator in the classical limit reads $\langle \hat{n}_0(\tau)\rangle = c_1 e^{-\tau\left(\gamma_Q + \sqrt{\gamma_Q^2-1}\right)} + c_2 e^{-\tau\left(\gamma_Q - \sqrt{\gamma_Q^2-1}\right)}$, and it is plotted in Fig. 2(a) of the main text together with the relative-number fluctuations.

### III. RELATIVE NUMBER FLUCTUATIONS

In order to derive analytically the relative number fluctuations we first consider the high temperature limit, where noise correlations become delta correlated in time. In particular, we assume $k_B T > \hbar\Omega_{\max}$, with $k_B$ being the Boltzmann constant and $\Omega_{\max}$ the maximum frequency that the quantum particle can adiabatically follow. Note that $\Omega_{\max}$ is smaller than the cut-off frequency defined by the Luttinger theory as otherwise we would thermally activate modes beyond the LL description [54]. Under this assumption and considering that the bath is in thermal equilibrium for which $\langle \xi_N(t)\rangle = 0$ we can approximate the hyperbolic cotangent term, $\coth(\hbar\Omega_\mu/(2k_B T))$, appearing in the correlations of the initial conditions of the bath modes $\langle \{\dot{\hat{q}}_\mu(t), \dot{\hat{q}}_{\mu'}(t')\}\rangle$, to first order and obtain:

$$\langle \xi_N(t)\xi_N(t')\rangle = \eta \delta(t-t') \tag{S11}$$

with $\eta = 2E_J^2 k_B T/\hbar^2 MLv$.

The time evolution for the number fluctuations is obtained using Eq. (S9) according to

$$\left\langle \hat{N}(\tau)^2\right\rangle = \langle \hat{n}_0(\tau)^2\rangle + 2\left\langle \hat{n}_0(\tau)\int_0^\tau dt'\, G_N(\tau,\tau')\chi_N(\tau')\right\rangle + \left\langle \int_0^\tau dt'\int_0^\tau d\tau''\, G_N(\tau,\tau')G_N(\tau,\tau'')\chi_N(\tau')\chi_N(\tau'')\right\rangle, \tag{S12}$$

with $\tau = \omega_0 t$. By introducing Eq. (S11) into the previous equation we find

$$\left\langle \hat{N}(\tau)^2\right\rangle = \langle \hat{n}_0(\tau)^2\rangle + \frac{ML^2\gamma_Q k_B T}{\hbar^2 \Omega^2}\left(\left[\frac{\Omega^2}{\gamma_Q(\gamma_Q^2+\Omega^2)}\right] - e^{-2\gamma_Q\tau}\left[\frac{1}{\gamma_Q} - \frac{\gamma_Q}{\gamma_Q^2+\Omega^2}\cos(2\Omega\tau) + \frac{\Omega}{\gamma_Q^2+\Omega^2}\sin(2\Omega\tau)\right]\right). \tag{S13}$$

Using the definition $\Delta N_{\rm LL} = \langle (\hat{N} - \langle \hat{N}\rangle)^2\rangle^{1/2}$ and taking the long-time limit, we readily obtain

$$\Delta N_{\rm LL} = \sqrt{\frac{ML^2}{\hbar^2}k_B T} \tag{S14}$$

It is interesting to compare the above result obtained within the classical limit with the full quantum mechanical calculation at equilibrium. Since at long times $\langle \hat{N}\rangle = 0$, it is sufficient to compute the symmetrized version of $\langle \hat{N}^2\rangle$, which readily follows from the fluctuation-dissipation theorem

$$\langle \{\hat{N}(0), \hat{N}(0)\}\rangle/2 = \int_{-\infty}^{+\infty} d\omega\, {\rm Im}[\tilde{G}_N(\omega)]\coth(\hbar\beta\omega/2)$$

$$= \frac{k_B T}{M_N L^2}\sum_{n=-\infty}^{n=+\infty}\frac{1}{\omega_0^2 + \nu_n^2 + |\nu_n|\gamma_0} = \frac{k_B T}{M_N L^2 \omega_0^2} +$$

$$\frac{\hbar}{M_N L^2 \pi(\lambda_2-\lambda_1)}(\psi(1+\lambda_2/\nu_1) - \psi(1+\lambda_1/\nu_1)) \tag{S15}$$

where $\nu_n = 2\pi n/\hbar\beta$ with $\beta = 1/k_B T$ are the Matsubara frequencies, $M_N = \hbar^2/E_J L^2$ is the effective mass, $\lambda_{1/2} = \gamma_0 \pm i\Omega_\lambda$ with $\Omega_\lambda = \sqrt{\omega_0^2 - \gamma_0^2}$ are the poles of the Fourier transform of the harmonic oscillator response function $\tilde{G}_N(\omega)$, $\psi(1 + \lambda_{1/2}/\nu_1)$ is the digamma function and $\{,\}$ are the Poisson brackets [60]. The first term in the right hand side of Eq. (S15), which is the leading order in the high-temperature limit, coincides with Eq.(S14). This indicates that the Josephson oscillations are damped towards the equilibrium state, where thermalization is provided by the phonon bath [38,39]. The subleading corrections in Eq. (S15) correspond to the contribution of the quantum fluctuations, which will become relevant at lower temperatures.

## IV. ADDITIONAL MATERIAL FOR THE EXACT SOLUTION IN THE TONKS-GIRARDEAU REGIME

### A. Details of the quench protocol

We use the time-dependent Bose-Fermi mapping [62-64] to describe the quench dynamics in the Tonks-Girardeau (TG) regime. To induce Josephson oscillations, we perform a quench in the confining potential $V(x,t)$ which excites the system creating a relative particle imbalance. In particular we use $V(x,t) = V_1(x)$ for $t \leq 0$ and $V_2(x)$ for $t > 0$ with $V_1(x) = U_0\delta(x) + \delta V_0 H(x)$, with $H(x)$ being the Heaviside step function, and $V_2(x) = U_0\delta(x)$ for $|x| \leq L$. At all times we assume hard-wall boundary conditions for $x = \pm L$.

The dynamics is then calculated by projecting the initial state onto the after-quench eigenbasis $\chi_\ell(x)$ with eigenvalues $\epsilon_\ell$, i.e.,:

$$\psi_n(t \leq 0) = \psi_n, \qquad \psi_n(t > 0) = \sum_{\ell}^{\infty} \langle \chi_\ell | \psi_n \rangle \chi_\ell(x) e^{-i\epsilon_\ell t/\hbar}, \tag{S16}$$

where $\psi_n(x)$ are the eigenstates of the pre-quench Hamiltonian.

After the quench, $2N_{\text{ex}}$ excitations are produced, each one with their corresponding associated frequencies. These excitations lead to a relative particle dynamics in which the oscillation frequency $\omega_{TG}$ and an envelope frequency $\omega_{\text{env}}$, at zero temperature and even number of particles read: $\omega_{TG} = \frac{1}{2N_{\text{ex}}} \sum_{n=-N_{\text{ex}}+1}^{N_{\text{ex}}} \omega_{N+2n}$ and $\omega_{\text{env}} = \frac{1}{2} \sum_{n=-N_{\text{ex}}+1}^{N_{\text{ex}}} (-1)^n \omega_{N+2n}$ with $\omega_n = \epsilon_n - \epsilon_{n-1}$ being the energy difference between the $n$-th and $(n-1)$-th single-particle energy levels of the after-quench system.

For odd number of particles, excitations are created in pairs due to the small gap between adjacent even odd states and an analogous expression can be obtained for the frequencies of the envelope and of the fast oscillations.

### B. Linear response theory

In this section we provide the details of the linear-response (LR) theory for the Tonks-Girardeau gas, used in the main text to compare the TG results with the Luttinger liquid ones.

The linear response theory describes the dynamics induced by a small space-time perturbation, which in our case is the step potential, i.e. we take $V(x,t) = \delta V_0 H(t) H(x)$, with $H(x)$ the Heaviside function. Within linear response theory the evolution of the density fluctuation is given by:

$$\langle \hat{n}(x,t) \rangle = \int_{-\infty}^{\infty} dx' \int_{-\infty}^{\infty} dt' \chi(x;x';t-t') V(x',t'), \tag{S17}$$

with $\langle \hat{n}(x,t) \rangle = \langle \hat{\rho}(x,t) \rangle - \rho_0(x)$ with $\rho_0(x)$ being the equilibrium density and $\chi(x;x';t-t') = (1/i\hbar)H(t-t') \langle [\hat{\rho}(x,t), \hat{\rho}(x',t')] \rangle$ the density-density response function. In the TG limit the density-density response function coincides with the one of a non-interacting Fermi gas and reads:

$$\chi(x;x';\omega) = \frac{1}{\hbar} \sum_{j \neq \ell} \psi_j^*(x)\psi_\ell(x)\psi_\ell^*(x')\psi_j(x') f(\epsilon_j)[1 - f(\epsilon_\ell)]$$
$$\times \left( \frac{1}{(\omega - (\epsilon_\ell - \epsilon_j)/\hbar) + i0^+} - \frac{1}{(\omega + (\epsilon_\ell - \epsilon_j)/\hbar) + i0^+} \right) \tag{S18}$$

where $\epsilon_j$ and $\psi_j(x)$ are the energies and single-particle orbitals of the unperturbed Hamiltonian and $f(\epsilon) = 1/[\exp((\epsilon - \mu)/k_B T) + 1]$ is the Fermi-Dirac distribution, with $\mu$ being the fermionic chemical potential.

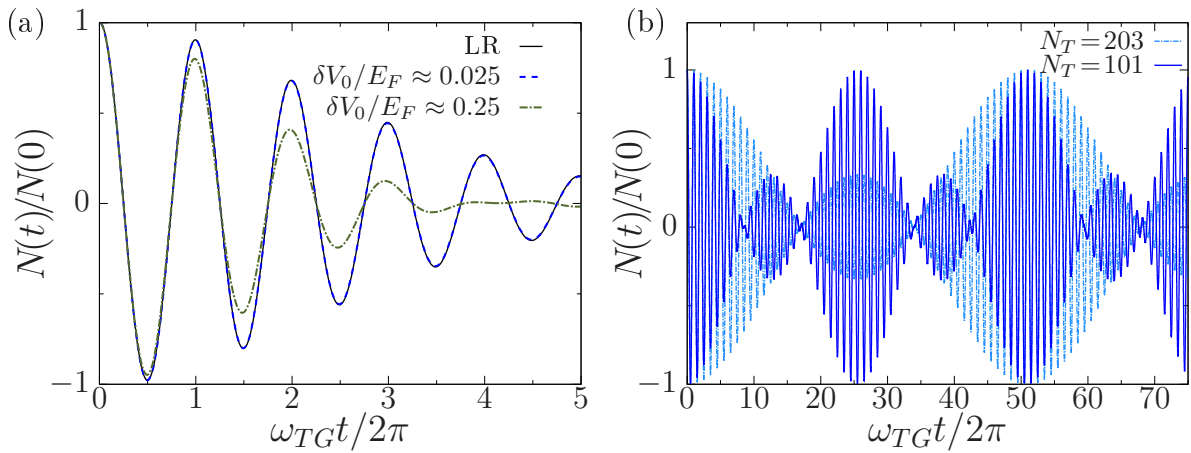

FIG. 1. (a) Dynamics of the relative-particle imbalance in the TG regime: LR approach (solid line) and the exact quench results (dashed, and dot-dashed lines, for two different values of $\delta V_0$ as indicated on the figure). The calculations are performed at finite temperature, with $T/T_F = 1/25$, dimensionless barrier strength $\lambda = 2mLU_0/(\hbar^2\pi^2 N) = 200$ and total number of particles $N_T = 101$. (b) TG long-time dynamics from the exact quench results at zero temperature for number of excitations $N_{\rm ex} = 3/2$ and two choices of the total number of particles $N_T = 101$ (dashed-blue line) and $N_T = 203$ (solid-dark-blue line).

By writing the step potential in Fourier space as $V(x,\omega) = \delta V_0 H(x) \left(\pi\delta(\omega) + i\frac{P}{\omega}\right)$ with $P$ indicating the principal part over the frequency domain, and by performing a contour integral on the lower part of the complex plane, we can rewrite Eq. (S17) as

$$\langle \hat{n}(x,t)\rangle = \frac{2\delta V_0}{\hbar} \sum_{j\neq\ell} \frac{\cos(\omega_{j,\ell}t)}{\omega_{j,\ell}} f(\epsilon_j)[1 - f(\epsilon_\ell)]\psi_j^*(x)\psi_\ell(x) \int_0^L dx'\psi_\ell^*(x')\psi_j(x'), \tag{S19}$$

with $\omega_{j,\ell} = \epsilon_\ell - \epsilon_j$. The dynamical evolution of the relative-number is then obtained as $N(t) = \int_{-L}^0 dx\, \langle \hat{n}(x,t)\rangle - \int_0^L dx\, \langle \hat{n}(x,t)\rangle$

$$\langle N(t)\rangle = \frac{4\delta V_0}{\hbar} \sum_{j\neq\ell} \cos(\omega_{j,\ell}t)f(\epsilon_j)[1 - f(\epsilon_\ell)]\frac{A_-(j,\ell)A_+(j,\ell)}{\omega_{j,\ell}}, \tag{S20}$$

with $A_-(j,\ell) = \int_{-L}^0 dx\psi_j^*(x)\psi_\ell(x)$, $A_+(j,\ell) = \int_0^L dx'\psi_\ell^*(x')\psi_j(x')$

Figure 1(a) of this supplemental material shows the relative-number oscillations obtained within the linear response theory as compared to the predictions of the exact quench solution. For small values of the step potential amplitude the agreement is excellent, thus validating the LR approach. For large values of $\delta V_0$ the exact dynamics is more damped than the LR solution, and the dynamics cannot be described in such a simplified picture.

In order to illustrate the type of excitations involved in the dynamics of particle imbalance, we restrict then to the linear response regime. In this case one can represent the excitations contributing to Eq. (S20), which are particle-hole oscillations, with the aid of the spectral function

$$S_{exc}^{TG}(k,\omega) = \sum_{j\neq\ell} \delta_{\omega,\omega_{j,\ell}}\delta_{k,k_\ell - k_j}f(\epsilon_j)[1 - f(\epsilon_\ell)], \tag{S21}$$

where $\delta_{\alpha,\beta}$ is the Kroenecker delta function. The spectral function is illustrated in Fig.3(c) of the main paper. Note that wave-vectors and energies are discrete as a consequence of the finite size of the wires considered in this work. Then, as discussed in the main text, the excitation spectrum in the linear-reponse regime already allows to pinpoint the limits of validity of the Luttinger liquid theory.

### C. Long-time dynamics

We provide in this section an analysis of the long-dynamics of the exact quench solution of the TG regime at zero temperature. The results are illustrated in Fig. 1(b) for two choices of total number of particles. We find that revivals appear in the dynamics, as a result of resynchronization of the different excitations created by the quench of the step potential. We also notice that the revival time increases with the total number of particles.

\clearpage

\end{document}